\documentclass[a4paper]{jpconf}

\usepackage{graphicx}

\begin{document}
\title{Phenomenological studies of double charged pion electroproduction from
the CLAS data.}

\author{Victor Mokeev, Volker Burkert.}

\address{Thomas Jefferson National Accelerator Facility, Newport
News, Virginia 23606, USA.}

\ead{mokeev@jlab.org}

\begin{abstract}
Recent results from phenomenological studies of the CLAS data on 2$\pi$
electroproduction off proton are presented. The analysis is focused on 
extracting of N $\rightarrow$ $N^{*}$ electromagnetic transition amplitudes from
the full data set on unpolarized 1-differential 2$\pi$ cross-sections.
\end{abstract}

\section{Introduction}

The data from  $\gamma p \rightarrow \pi^{-} \pi^{+} p$ exclusive processes
 play an important role in
studies of nucleon resonances with the CLAS detector \cite{Bu05}. 
Single and double pion channels combined account for major part of the total 
cross-section in the $N^{*}$ excitation region. These two channels offer complementary
information on $N^{*}$'s. The single pion channel is mostly sensitive to 
resonances with
masses less than 1.6 GeV, while many high lying states preferably decay
with 2$\pi$ emission. The studies of 2$\pi$ electroproduction with CLAS for the
first time provided the electrocouplings of high lying nucleon states
\cite{jm06,pa06}. Moreover, 1$\pi$ and 2$\pi$ channels are strongly coupled 
by final state interactions (FSI).
Therefore, data on 2$\pi$ electroproduction are critical for $N^{*}$
studies in the combined analysis of major exclusive channels within the framework
of coupled channel approaches. The most advanced coupled channel analysis,
incorporating rigorous treatment of FSI
in the 3-body $\pi\pi N$ final states was proposed recently \cite{Lee06}.

For the first time a complete set of unpolarized
single-differential cross sections was measured
\cite{Ri03,Mo06}.
Phenomenological analysis of these data 
open up the opportunity to establish the major contributing mechanisms and to
determine the evolution of $N^{*}$ electrocouplings in a wide range of photon
virtualities.

\section{Two-pion electroproduction mechanisms from phenomenological
analysis.}

We have developed a phenomenological model, that incorporates particular 
meson-baryon mechanisms
based on their manifestations in observables$:$ as enhancements in invariant
mass distributions, sharp forward/backward slopes in angular distributions.
Analysis of earlier CLAS data \cite{Ri03} incorporated
 particular
meson-baryon mechanisms needed to describe $\pi^{+}$p, $\pi^{+}$$\pi^{-}$,
$\pi^{-}$p invariant masses and
$\pi^{-}$ angular distribution. These cross sections were analyzed in the
hadronic mass range from 1.41 to 1.89 GeV. The overall $Q^{2}$-coverage ranges 
from 0.5 to 1.5 $GeV^{2}$. In the 2005 version of this analysis approach 
(JM05) \cite{jm06}
double charged pion 
production was described by the 
superposition of quasi-two-body channels with the formation and
subsequent decay of unstable particles in the intermediate states: 
$\pi^{-} \Delta^{++}$, $\pi^{+} \Delta^{0}$, $\rho^{0} p$, 
$\pi^{+} D^0_{13}(1520)$, $\pi^{+} F^0_{15}(1520)$, $\pi^{-} P^{++}_{33}(1640)$.
Remaining  direct 2$\pi$ production mechanisms without formation of unstable
intermediate particles were described by a set of exchange terms with the
amplitudes as outlined in \cite{Az05}.  
Implementation of exchange terms allowed much improved description
of the $\pi^{-}$ angular distributions \cite{Mo05c}.

The production amplitudes for the first three quasi-two-body 
intermediate states were treated as sums
of $N^{*}$ excitations in the $s$-channel and non-resonant mechanisms
described in Refs \cite{jm06,Mo01}. All well established 4 star resonances
with observed decays to the two pion
final states were included as well as the 3-star states
$D_{13}(1700)$, $P_{11}(1710)$, $P_{33}(1600)$, and $P_{33}(1920)$.

The production amplitudes for the $\pi^{+} D^0_{13}(1520)$, 
$\pi^{+} F^0_{15}(1520)$, $\pi^{-} P^{++}_{33}(1640)$ intermediate state were
treated as non-resonant contributions only. Their parametrization is described
in \cite{jm06,Mo05c}.

In the JM05 approach we succeeded in describing all
beforementioned observables in the CLAS data \cite{Ri03}. 
These results are presented 
in \cite{Mo05c}. 

\section{Global analysis of unpolarized cross-sections in 2$\pi$ electroproduction}.

\begin{figure}[htbp]
\vspace{5.0cm}
\includegraphics{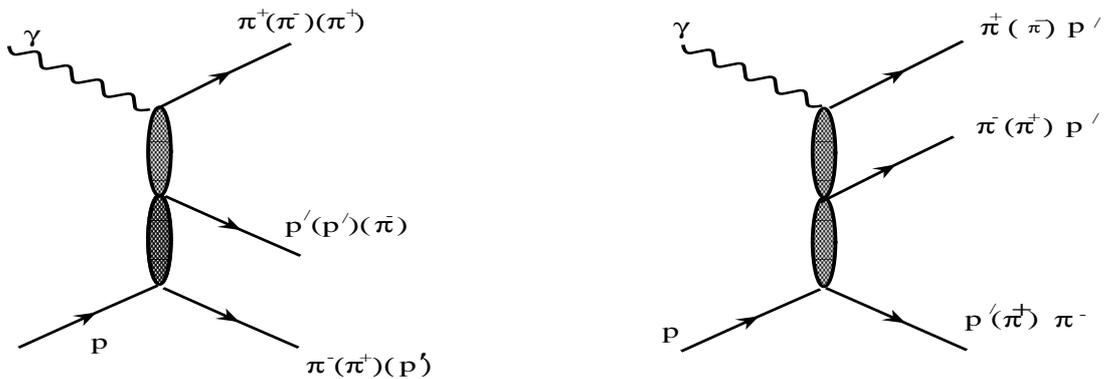} 
\caption[]{\small{Direct 2$\pi$ production mechanisms in JM06 model}}
\label{diag}
\end{figure}

In the analysis of preliminary CLAS 2$\pi$ data at W$<$1.6 GeV and photon
virtualities from 0.2 to 0.6 $GeV^{2}$ \cite{Mo06} for the first time we
attempted to
fit contributing mechanisms to the full set of unpolarized
single-differential cross-sections. Nine single-differential cross-sections combined were 
fitted within the framework of JM05 approach 
 in each
W and  $Q^{2}$ bin covered by measurements \cite{Mo06}. 
They consist of all
single-differential cross-sections mentioned in previous section and also included
 $\pi^+$ and p 
angular distributions
and three distributions over angles $\alpha_{i,j}$ between two planes, composed 
by momenta of
the two pairs of the
final hadrons for three combinations amongst these pairs. To provide
a better description of $\pi^+$ and p angular distributions, we modified the
dynamics of direct 2$\pi$ production mechanisms with respect to those 
used in JM05
version. 
The mechanisms of \cite{Az05} were substituted by ladder-type double 
exchange processes, shown in Fig.~\ref{diag}

\begin{figure}[htbp]
\vspace{11.0cm}
\includegraphics{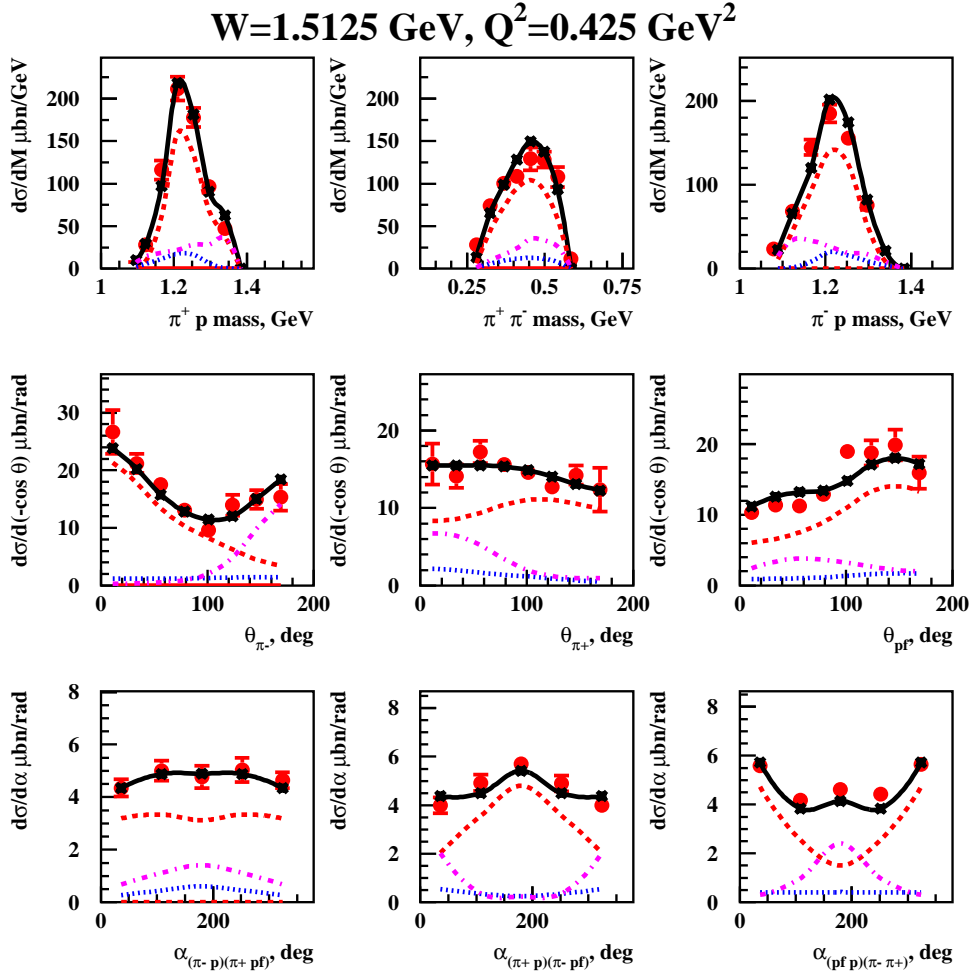} 
\caption[]{\small{Complete set of unpolarized 1-differential cross-sections
at W=1.51 GeV and $Q^{2}$=0.425 $GeV^{2}$} in comparison with JM06 results
(solid lines). The contributions from $\pi^{-} \Delta^{++}$, $\pi^{+}
\Delta^{0}$ isobar channels and direct 2$\pi$ production mechanisms are shown 
by dashed, dotted  and dot-dashed lines respectively.  }
\label{1diff}
\end{figure}

 The amplitude parametrization is
presented in \cite{Mo06}. After these modifications we succeeded
to describe CLAS data \cite{Mo06} in the entire kinematics covered by
the measurements. As a typical example, description of an entire set of 
single-differential 2$\pi$
cross-sections within the framework of the JM06 model version is shown in 
Fig.~\ref{1diff} together with the contributions
from various mechanisms of JM06 approach.

The shapes of cross-sections for various contributing mechanisms are
substantially
different in the observables, but highly correlated by mechanism
amplitudes. Therefore, a successful description of all unpolarized 
single-differential 
cross-sections allowed us to pin down all major
contributing processes. We found no need for 
remaining mechanisms of unknown dynamics. To check the reliability of 
the amplitudes for contributing
processes, derived in phenomenological data analysis, we fixed all
JM06 parameters, fitting them to  six single-differential cross-sections$:$ all invariant
masses and three final state angular distributions. The remaining three 
distributions 
over $\alpha_{i,j}$'s angles were calculated, keeping JM06 parameters fixed. 
Reasonable description of $\alpha_{i,j}$'s angular
distributions was achieved in the entire kinematics area covered 
by measurements. Therefore, we
confirmed the reliability of 2$\pi$ electroproduction mechanisms established 
in phenomenological data
analysis within the framework of JM06 model.

\begin{figure}[htbp]
\vspace{9.0cm}
\includegraphics{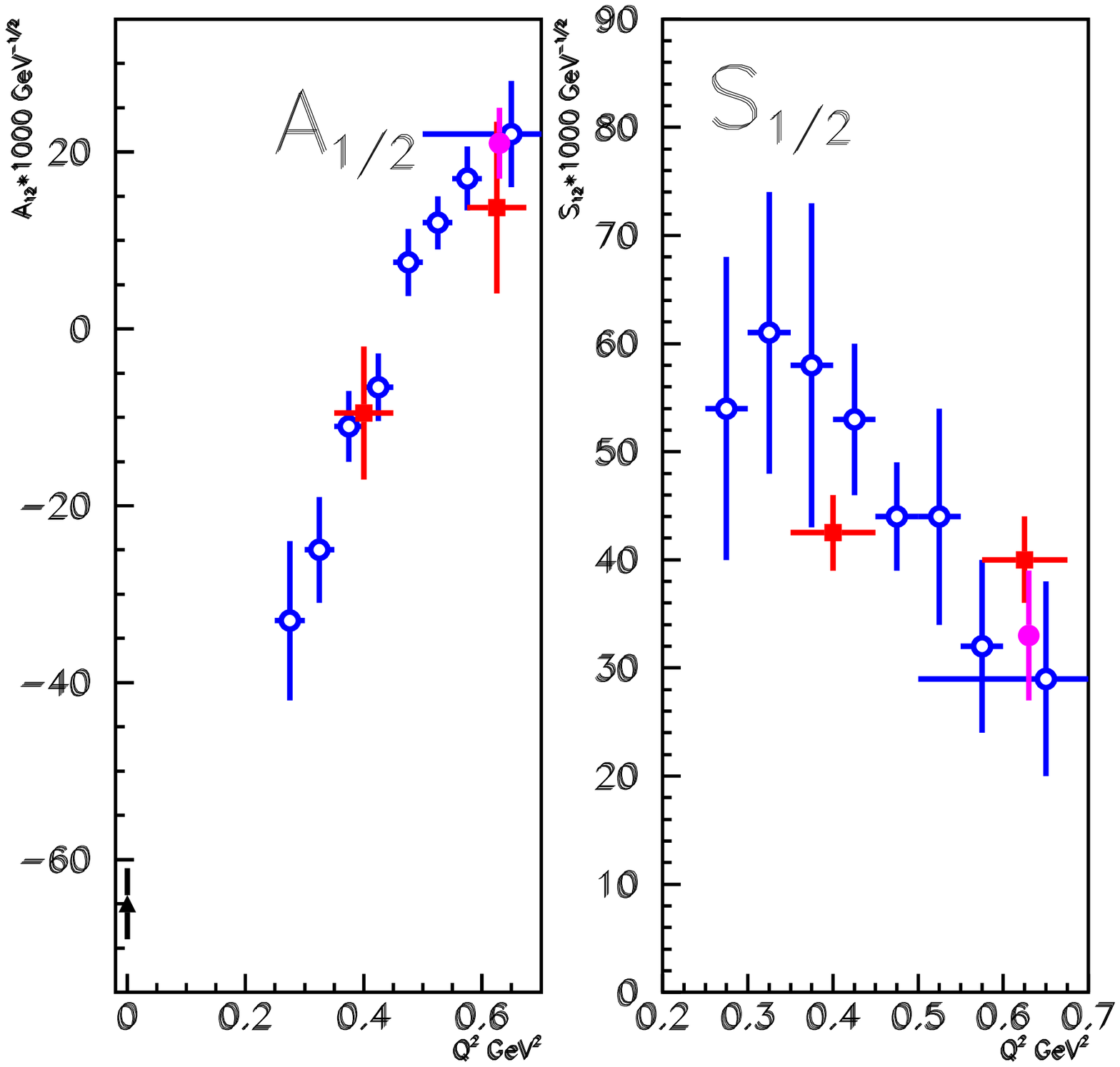} 
\includegraphics{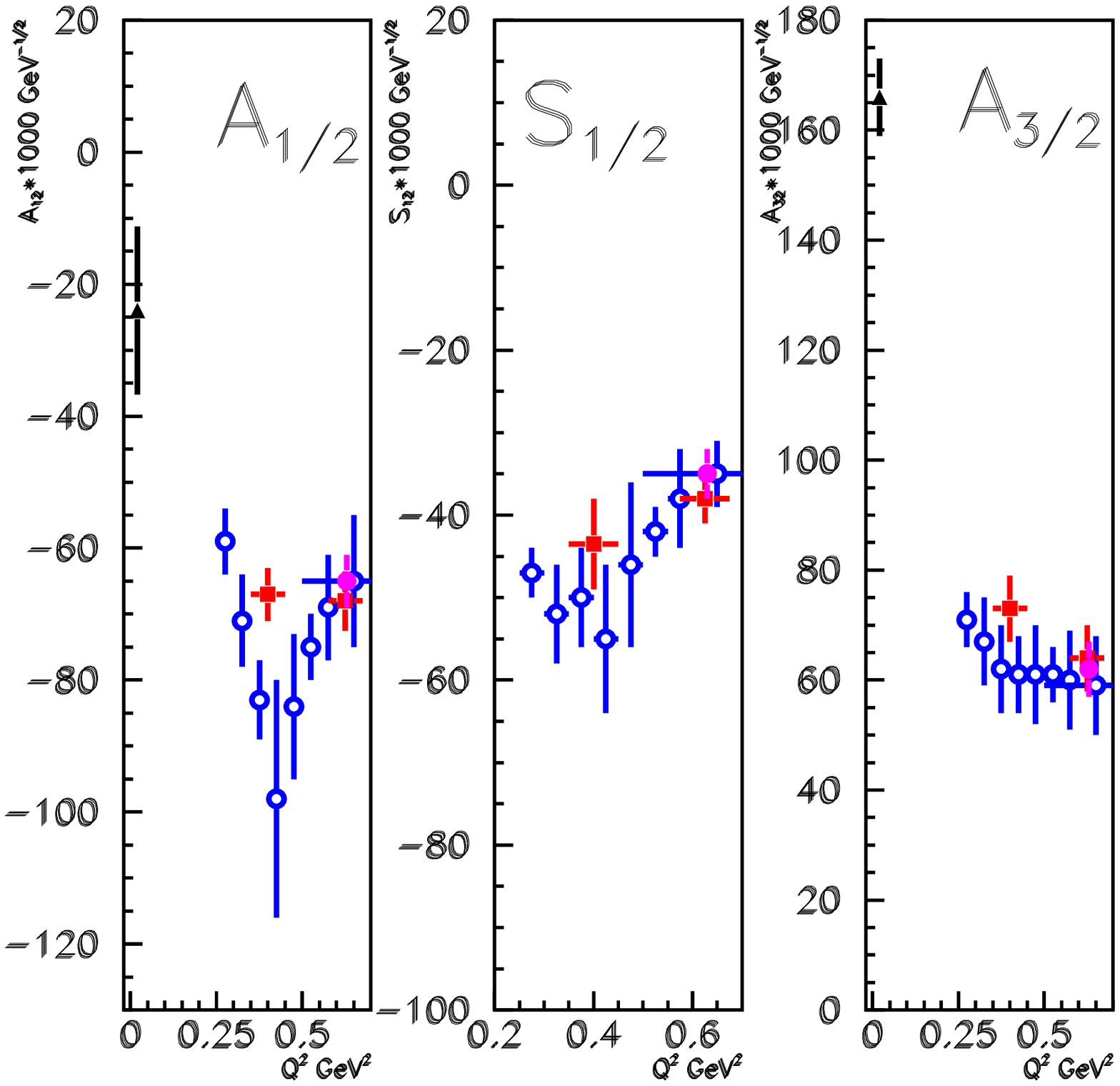} 
\caption[]{\small{Electromagnetic transition form factors for the $P_{11}(1440)$ (left) and  
$D_{13}(1520)$ (right) states, determined from analysis of CLAS single pion data 
(squares) and double pion data (open circles). The results from 
the combined 
1$\pi$/2$\pi$ data analysis at $Q^2$=0.65~GeV$^2$ are shown by filled
circles. The 
photocouplings from the PDG are shown by black triangles.}}
\label{p11d13}
\end{figure}

In Fig.~\ref{p11d13} we show electrocouplings of $P_{11}(1440)$, $D_{13}(1520)$ 
states, determined from the analysis of preliminary CLAS data on 2$\pi$
electroproduction \cite{Mo06} within the framework of JM06 approach. For the 
first time we
obtained information on $Q^{2}$-evolution of electrocouplings for these
states from $\pi^{-}\pi^{+} p$ channel at
$Q^{2}$ from 0.2 to 0.6 $GeV^{2}$. These photon virtualities are particularly
sensitive to the contributions from $N^{*}$ meson-baryon dressing. Moreover,
 these data are not contaminated by the contribution from $P_{33}(1232)$ tail.
 The
electrocouplings
obtained from this analysis are in reasonable agreement
with the results from 1$\pi$ exclusive channel \cite{Az05b}, as well as from 
the combined
1$\pi$/2$\pi$ analysis \cite{Az05}. 

\section{Conclusions}.

A phenomenological model  (JM06) has been developed for the description of 2$\pi$
electroproduction in $N^{*}$ excitation region with the most complete accounting
for contributing mechanisms. Successful description of all unpolarized 
single-differential cross-sections was achieved at W$<$1.6 GeV and at photon virtualities
from 0.2 to 0.6 $GeV^2$. Electrocouplings of $P_{11}(1440)$, $D_{13}(1520)$
states were derived from the combine fit of all unpolarized
observables in 2$\pi$ electroproduction.

$\ $



\begin{thebibliography}{99}

\bibitem{Bu05} 
V.D. Burkert, {\it Prog. Part. Nucl. Phys.} {\bf 55}, 108 (2005).



\bibitem{jm06} 
V.I. Mokeev, V.D. Burkert {\it et al.}, {\it Proceedings of the Workshop on Physics 
of Excited Nucleons NSTAR2005}, ed. by S.Capstick, V.Crede, P.Eugenio, p.47-56.



\bibitem{pa06} 
V.I. Mokeev, V.D. Burkert, {\it AIP Conf.} Proc. {\bf 842}, 339 (2006). 





\bibitem{Lee06} 
A. Matsuyama, T. Sato, and T.-S. H. Lee, ``{\it Dynamical coupled channel
model of meson production reactions in nucleon resonance region}'', 
nucl-th/0608051, (2006).

\bibitem{Ri03} 
M. Ripani {\it et al.}, {\it Phys. Rev. Lett.} {\bf 91}, 022002 (2003), see
full data set in CLAS Physics Data Base, $http://clasdb3.jlab.org$.



\bibitem{Mo06} 
V.I. Mokeev, Talk on N* Analysis Workshop, November 4-6 2006, Newport News, VA, 
http://conferences.jlab.org/Nstar/program.html.


\bibitem{Mo01} V. Mokeev et. al.,
{\it Phys. of Atom. Nucl.}  {\bf 64}, 1292 (2001).



\bibitem{Az05} 
I.G. Aznauryan {\it et al.}, {\it Phys. Rev.} C {\bf 72}, 045201 (2005).




\bibitem{Az05b} 
I.G. Aznauryan {\it et al.}, {\it Phys. Rev.} C {\bf 71}, 015201 (2005). 




\bibitem{Mo05c} V.I.Mokeev $http://hadron.physics.fsu.edu/nstar/scientificProg.htm$


 

 

\end{thebibliography}
\end{document}